\documentclass[12pt]{article}
\usepackage{epsf}
\usepackage{a4wide,graphicx}
\usepackage{cite}
\usepackage{subfigure}

\newcommand{\beq}{\begin{equation}}
\newcommand{\eeq}[1]{\label{#1} \end{equation}}

\title{Proton emission off nuclei induced by kaons in flight}

\author{
 V. K. Magas$^1$, J. Yamagata-Sekihara$^{2,3}$, S. Hirenzaki$^{4}$, E. Oset$^{3}$,\\
 A. Ramos$^{1}$  \\
{\small{\it $^1$Departament d'Estructura i Constituents de la Materia, Universitat de Barcelona,}}\\
{\small{\it  Diagonal 647, E-08028 Barcelona, Spain}}\\
{\small{\it $^2$Yukawa Institute for Theoretical Physics, Kyoto University, Kyoto 606-8502, Japan}}\\
{\small{\it $^3$ Departamento de F\'{\i}sica Te\'orica and IFIC, Centro Mixto Universidad de Valencia-CSIC,}}\\
{\small{\it Institutos de Investigaci\'on de Paterna, Apartado 22085, 46071 Valencia, Spain}}\\
{\small{\it $^4$ Department of Physics, Nara Women's University, Nara 630-8506, Japan}}\\
}
\date{\today}
\begin{document}
\maketitle

\begin{abstract}

We study the $(K^-,p)$ reaction on nuclei with a 1 GeV/c momentum kaon beam,
paying a special attention at the region of emitted protons having kinetic
energy above 600 MeV, which was used to claim a deeply attractive
kaon nucleus optical
potential. Our model describes the nuclear reaction in the framework of
a local density
approach and the calculations are performed following two different
procedures:
one is based on a many-body method using the Lindhard function and the
other one is based on a Monte Carlo simulation. The simulation
method offers flexibility to account for
processes other than kaon quasi-elastic scattering, like
$K^-$ absorption by one and two nucleons producing hyperons, and allows
to consider final state interactions of the $K^-$, $p$ and all other
primary and
secondary particles on their way out of the nucleus, as well as the weak
decay of the produced hyperons into $\pi N$.
We find a limited sensitivity of the cross section to the strength of
the kaon optical potential. We also show serious drawback in the
experimental set up from the requirement of having, together with
 the energetic proton, at least one charged particle
detected in the decay counter surrounding the target, since we find that
the shape of the original cross section is appreciably distorted, to the
point of invalidating the claims made in the experimental paper on the
strength of the kaon nucleus optical.

\end{abstract}

\section{Introduction}

     The issue of the kaon interaction in the nucleus has attracted much
attention in past years. Although from the study of kaonic atoms one knows that
the $K^-$-nucleus potential is attractive \cite{friedman-gal}, the discussion
centers on how attractive the potential is and if it can accommodate deeply
bound kaonic atoms (kaonic nuclei), which could be observed in direct reactions.
 A sufficiently large attraction could even make possible the existence of
 kaon condensates in  nuclei, which has been suggested in \cite{Kaplan:1986yq}.
 Stimulated by the success in reproducing the data of kaonic 
atoms, many
works considered strongly attractive potentials of the order of 200 MeV
at normal nuclear matter density
\cite{gal1,gal,muto,amigo1,amigo2}, or explored the dependence of a few
observables to a wide range of depths from 0 to 200 MeV
\cite{gal2,gal3,gal4}.
More moderate attraction is found in similar works done in
\cite{Zhong:2004wa,Zhong:2006hd,Dang:2007ai}. 
Yet, all modern potentials
based on underlying chiral dynamics of the Kaon-nucleon interaction
\cite{lutz,angelsself,schaffner,galself,Tolos:2006ny} lead to
moderate potentials of the order of 60 MeV attraction at nuclear matter density.
They also
have a large imaginary part making the width of the deeply bound states
much larger than the energy separation between the levels, which would rule out
the  experimental observation of peaks. The agreement with the
data of kaonic atoms of this purely theoretical shallow potential is good
\cite{okumura}, and a fit to all data adding a small phenomenological potential
to the theoretical one performed in  \cite{baca} indicates that the best fit
potential  deviates at most by 20\% from the theoretical one of
\cite{angelsself}.

The opposite extreme is represented by some highly
    attractive phenomenological potentials with about
600 MeV strength in the center of the nucleus \cite{akaishi:2002bg,akainew}.  These
potentials, leading to compressed
nuclear matter of ten times nuclear matter density, met criticisms from
\cite{toki}  and more
recently from  \cite{Hyodo:2007jq}, which were rebutted in
\cite{akanuc} and followed by further argumentation in  \cite{Oset:2007vu} and
 \cite{npangels}. More recently the lightest K-nuclear system of $\bar{K}NN$
 has also been the subject of strong debate
\cite{shevchenko,hyodo,sato,akaishi:2007cs}.

Experimentally, the great excitement generated by peaks seen at KEK
\cite{Suzuki:2004ep} and FINUDA \cite{Agnello:2005qj,:2007ph}, originally
interpreted in terms of deeply bound kaons atoms, is receding, particularly
after the work of \cite{toki}, regarding the KEK experiment, and those of
\cite{Magas:2006fn,Crimea,Magas:2008bp}, regarding the FINUDA ones, found
explanations of the  experimental peaks based on conventional reactions that
unavoidably occur in the process of kaon absorption. Also the thoughts of
\cite{Suzuki:2007kn}, with opposite views to those of FINUDA in
\cite{:2007ph},
and the reanalysis of the KEK proton spectrum from $K^-$ absorption on
$^4$He
\cite{Suzuki:2004ep}, done in \cite{Sato:2007sb}, where the original narrow
peak appears much broader and is consistent with the signal seen on a
heavier
$^6$Li target in FINUDA \cite{Agnello:2006jt}, have
helped to bring the discussion to more realistic terms. Nevertheless, the
possibility that the FINUDA peak of  \cite{Agnello:2005qj} could be a
signal of
a deeply bound kaon state is still defended \cite{piano}. This brief
description
just shows the intense activity and strong interest in this subject over the
past few years.

There are also claims (with very low statistical significance) of
$K^-pp$ and
$K^-ppn$ bound states from $\bar{p}$ annihilation in $ ^4$He at rest
measured
by OBELIX@CERN \cite{Obelix}, as well as the recent claim of a $K^-pp$ bound
state, seen from the $pp\rightarrow K^+ X$ reaction by the DISTO experiment
\cite{DISTO}. These experimental claims are under investigation now since,
before calling in new physics, it is important to make clear that these data
cannot be explained with conventional mechanisms.

     In this work we focus on yet another experiment which led the authors to claim
evidence for a very strong kaon-nucleons potential, with a depth
of the order of 200 MeV \cite{Kishimoto:2007zz}.  The experiment
looks for fast protons
emitted from the absorption of in flight kaons by nuclei. Our aim is to
show how this experiment was analyzed and which ingredients are missing.
Throughout this work we shall see that the
interpretation of the data requires a more thorough analysis, and with
all things considered, we reach different conclusions than those of Ref.~\cite{Kishimoto:2007zz},
in the sense that we do not find evidence for a
a strongly attractive kaon-nucleus optical potential.

   One of the shortcomings of Ref.~\cite{Kishimoto:2007zz} stems from employing the
Green's function method \cite{Morimatsu:1994sx} in a variant used in
\cite{zaki,Yamagata:2007cp,YamagataSekihara:2008ji} to analyze the data and extract
from there the kaon optical potential.  The only mechanism considered in
Ref.~\cite{Kishimoto:2007zz} for the emission of fast protons is the $\bar{K} p \to
\bar{K} p$ process, taking into account the optical potential for the slow
kaon in the final state. However, there are other mechanisms that
contribute to generate fast protons, namely kaon
absorption by one nucleon, $K^- N \to \pi \Sigma$ or $K^- N \to \pi \Lambda$
followed by decay of the $\Sigma$ or the $\Lambda$ into $\pi N$, or the
absorption by pairs of nucleons, $\bar{K} N N\to  \Sigma N$
and $\bar{K} N N\to  \Lambda N$, followed also by similar hyperon decays.
The contributions from these processes were also suggested in Ref. \cite{YH}.
In the present work, we take
into account these additional mechanisms by means
 of a Monte Carlo simulation, while the processes involving $\bar{K}N$
scattering, which are dominant in this reaction, are considered in two
different ways, one based on standard many body
methods using Lindhard functions and another one based on a Monte Carlo simulation.
The agreement of the two calculational methods gives us confidence to use
the Monte Carlo simulation for the processes involving more than one step and/or
one nucleon and two nucleon kaon absorption.

\section{The $(K^-,p)$ reaction in nuclei: many body approach}
\label{many-body}

   We are dealing here with an inclusive reaction, where a kaon in flight hits
a nucleus and a proton is emitted.  In the present case we focus on fast
protons, which
could be emitted when the kaons are trapped in the nucleus or are rescattered
 with small
energy. The reaction is inclusive in the sense that, apart from the proton
observed, anything  can happen to the nucleus. In fact many processes may take
place. The original kaon can undergo quasi-elastic
collisions with the nucleons, transferring them some energy. The kaon can be
absorbed, either by one nucleon or by pairs of
nucleons. The kaon can be trapped in a kaonic orbit, etc. Once a kaon has
experienced a particular reaction, the final products also suffer their
own interactions
with the nucleus before, eventually, a fast proton gets out.
 Complicated as it may sound ---and we shall deal with these
complications in a following section--- the evaluation of the inclusive cross
section is however easy since one is looking for the most energetic
protons and in the forward direction, which means that
these protons must take the
largest possible energy from the original kaon. In other words,
if a
kaon undergoes a
quasi-elastic collision in which the proton does not fall
in this narrow window, the event will be dismissed because this kaon,
having lost a fraction of its energy, will not have a second chance of
producing  a fast proton again. Obviously, if the kaon is absorbed it
disappears from the flux
and must also be eliminated in the evaluation of further processes
(the contribution from the absorption mechanisms will be calculated later).
The other relevant observation is
that if the final fast proton has a secondary collision it will also loose energy
and will not lie in the desired energy window.  In
practical terms this means that we  can just care about the
direct $K^ - p \rightarrow K^- p $ quasi-elastic reaction at a certain point in the nucleus
and distort the initial
$K^-$ and final proton waves.  These type of reactions have received
much attention and we describe here
the standard procedure of dealing with them  \cite{wanda,electron}.

Diagrammatically, the process considered is depicted in Fig. 1, which shows the
kaon inducing a $ph$ excitation and remaining in the nucleus, representing
what we call a quasi-elastic collision. The kaon transfers energy and momentum
to a nucleon that is promoted from below to above the (local) Fermi sea.

\begin{figure}[htpd]
\begin{center}
\includegraphics[width=5cm]{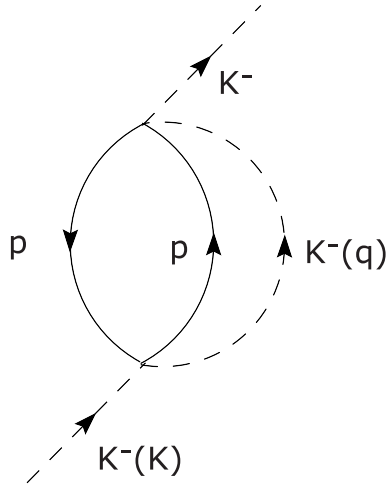}
\caption{\label{fig:1}
Diagrammatic representation of the inclusive ($K^-,p$) reaction.}
\end{center}
\end{figure}

The kaon self-energy for the diagram of Fig.~\ref{fig:1} in a Fermi sea is given by
\begin{equation}
-i\Pi_{\rm qe}(k)=\int \frac{d^4q}{(2\pi)^4}i{\bar U}(k-q)(-i)T(-i)T
 \frac{i}{{q^0}^2-{\vec q}\,^2-m_K^2-\Pi(q)}\,,
\label{eq:1}
\end{equation}
where ${\bar U}(k-q)$ is the Lindhard function for $ph$ excitation and $T$ stands for the $K^-p\to K^-p$ scattering matrix.
The factor $[{q^0}^2-{\vec q}\,^2-m_K^2-\Pi(q^0,{\vec q}\,)]^{-1}$ is the kaon
propagator which includes its self-energy in the medium, $\Pi(q^0,{\vec q}\,)$.
Using the Cutkosky rules one easily obtains the imaginary part of $\Pi_{\rm qe}(k)$ as in \cite{rafaloren},
\begin{eqnarray}
\Pi_{\rm qe}&\to&2i{\rm Im}~\Pi_{\rm qe}\,,\\
{\bar U}(k-q)&\to&2i\theta(k^0-q^0){\rm Im}~{\bar U}(k-q)\,,\\
D(q)= \frac{1}{{q^0}^2-{\vec q}\,^2-m_K^2-\Pi(q)}&\to&2i\theta(q^0){\rm Im}~D(q)~~.
\label{Eq:cuttheta}
\end{eqnarray}
Then
\begin{eqnarray}
{\rm Im}~\Pi_{\rm qe}(k)&=&-2\int\frac{d^4q}{(2\pi)^4}{\rm Im}~{\bar U}(k-q)|T|^2{\rm
Im}\frac{1}{{q^0}^2-{\vec q}\,^2-m_K^2-\Pi(q^0,{\vec q}\,)}\,,
\end{eqnarray}
where ${\bar U}$ is given by
\begin{equation}
{\bar U}(k-q)=2\int \frac{d^3 p_N}{(2\pi)^3}\frac{M}{E({\vec
p_N})}\frac{M}{E({\vec k}+{\vec p_N}-{\vec q}\,)}\frac{n({\vec p_N})[1-n({\vec
k}+{\vec p_N}-{\vec q}\,)]}{k^0+p_N^0-q^0-E_{N'}({\vec k}+{\vec p_N}-{\vec q}\,)+i\epsilon} \,.
\label{Eq:u}
\end{equation}

The fast protons with momentum ${\vec k}+{\vec p_N}-{\vec q}$ are
the energetic ones that would be observed, hence the corresponding
Pauli blocking factor, $1-n$, is just unity here.
>From Eq.~(\ref{Eq:u}) one obtains Im~${\bar U}$ as
\begin{equation}
{\rm Im}~{\bar U}=-2\pi\int\frac{d^3p_N}{(2\pi)^3}\frac{M}{E({\vec
p_N})}\frac{M}{E({\vec k}+{\vec p_N}-{\vec q}\,)}n({\vec p_N})\delta(k^0+E({\vec
p_N})-q^0-\Delta-E({\vec k}+{\vec p_N}-{\vec q}\,))\,,
\end{equation}
where we have introduced an energy gap between the energy of the holes and the energy of the particles \cite{Oset:1993sm,wambach}.
We thus obtain for Im~$\Pi_{\rm qe}$ the following equation
\begin{eqnarray}
{\rm Im}~\Pi_{\rm qe}(k)&=&2\int\frac{d^3p_N}{(2\pi)^3}n({\vec p_N})\frac{M}{E({\vec
p_N})}\int\frac{d^3q}{(2\pi)^3}|T|^2\frac{M}{E({\vec k}+{\vec p_N}-{\vec q}\,)}\nonumber\\
&&\times{\rm Im}\frac{1}{{q^0}^2-{\vec q}\,^2-m_K^2-\Pi(q^0,{\vec
q}\,)}\biggl|_{q^0=k^0+E({\vec p_N})-\Delta-E({\vec k}+{\vec p_N}-{\vec q}\,)}\,.
\end{eqnarray}
The physical interpretation comes by recalling that
\begin{eqnarray}
2\omega V_{\rm opt}&\equiv&\Pi_{\rm qe}\,,\\
{\rm Im}~V_{\rm opt}&=&\frac{1}{2\omega}{\rm Im}~\Pi_{\rm qe}\,,\\
\Gamma&=&-2{\rm Im}~V_{\rm opt}=-\frac{{\rm Im}~\Pi_{\rm qe}}{\omega}\,,
\end{eqnarray}
where $\omega$ is the kaon energy and $V_{\rm opt}$ is the $K^-$-nucleus optical potential.
Our states are normalized to unity in a box of volume $V$.
The flux of the incoming kaons is $v_{K^-}/V$ and thus the $K^-$ cross section with the nucleons of the Fermi sea of volume $V$ is given by
\begin{equation}
\sigma=\frac{\Gamma}{K^-~{\rm flux}}=\frac{\Gamma}{v_{K^-}/V}=V\frac{\Gamma}{k}\omega\,.
\end{equation}
Replacing $V$ by an integral $\int d^3r$ over the nuclear density we obtain, upon the change of variables
\begin{equation}
{\vec p}\,\equiv{\vec k}+{\vec p_N}-{\vec q} \, ,
\end{equation}
where ${\vec p}$ is the outgoing proton variable,
\begin{eqnarray}
\sigma&=&-\frac{2}{k}\int d^3r\int\frac{d^3p_N}{(2\pi)^3}n({\vec p_N},{\vec r}\,)\frac{M}{E({\vec p_N})}\nonumber\\
&&\times\int\frac{d^3p}{(2\pi)^3}|T|^2\frac{M}{E({\vec p}\,)}{\rm
Im}\frac{1}{{q^0}^2-{\vec q}\,^2-m_K^2-\Pi(q^0,{\vec
q}\,)}\biggl|^{q^0=k^0+E({\vec p_N})-\Delta-E({\vec p}\,)}_{{\vec q}\,={\vec
k}+{\vec p_N}-{\vec p}} \, ,
\end{eqnarray}
from where
\begin{eqnarray}
\frac{d\sigma}{d\Omega({\hat p})E({\vec p}\,)}&=&-\frac{2}{k}pM\int d^3r\int\frac{d^3p_N}{(2\pi)^3}n({\vec p_N},{\vec r}\,)\frac{M}{E({\vec p_N})}\nonumber\\
&&\times\frac{1}{(2\pi)^3}|T|^2{\rm Im}\frac{1}{{q^0}^2-{\vec
q}\,^2-m_K^2-\Pi(q^0,{\vec q}\,)}\biggl|^{q^0=k^0+E({\vec p_N})-\Delta-E({\vec
p}\,)}_{{\vec q}\,={\vec k}+{\vec p_N}-{\vec p}} \ .
\label{eq:dsigma}
\end{eqnarray}
The integral over ${\vec r}$ covers the size of the nucleus.
As in previous works \cite{Magas:2006fn,Crimea}, we use a realistic nuclear density profile for $^{12}$C, given by three-parameter Fermi distribution \cite{DeJager:1974dg}, 
which reproduces elastic electron scattering data.
We have also tried a more sophisticated nuclear density profile, which
accounts for the finite range of the interaction via a folding procedure,
and is preferred by the antiprotonic X-rays and radiochemical data \cite{folded}.
Although the folded density is almost $10\%$ lower in the center of nuclei and
extends to larger distances, the final result of our simulation is
practically unaffected by this density change.
The integral over ${\vec p}_N$ is restricted to the hole (bound) nucleon states within
the local Fermi momentum $k_F({\vec r}\,)$ obtained from the nuclear density at
point ${\vec r}$. This is accounted for by the Pauli blocking factor
$n(\vec{p}_N,\vec{r}\,)$.

In free space, the cross section for kaon scattering
off a proton in the lab frame with the proton emerging in the
forward direction reads
\begin{equation}
\frac{d\sigma}{d\Omega({\hat p})}\biggl|_{\rm lab}=\frac{\pi}{k}
\frac{1}{(2\pi)^3}{\bar p}^2|T|^2\frac{M}{2}
\frac{1}{{\bar p}(k^0+M)-E({\bar p})k} \ ,
\label{eq:free}
\end{equation}
where ${\bar p}$ is the momentum of the nucleon
\begin{equation}
{\bar p}=\frac{2p_{\rm CM}E_{\rm CM}}{M},
~~p_{\rm CM}=\frac{\lambda^{1/2}(s,m_K^2,M^2)}{2\sqrt{s}} \ .
\end{equation}
Equation (\ref{eq:free}) establishes a link between $|T|^2$ and the forward cross
section, which can be implemented into Eq.~(\ref{eq:dsigma}) to derive our final formula
$$
\frac{d\sigma}{d\Omega({\hat p})E({\vec p\,})}=-\frac{4p}{{\bar p}^2}\frac{d\sigma}{d\Omega({\hat p})}\biggl|_{\rm lab}\int d^3re^{-\int^\infty_{-\infty}\sigma\rho(b,z')dz'}\int\frac{d^3p_N}{(2\pi)^3}n({\vec p_N},{\vec r}\,)\frac{M}{E({\vec p_N})}\theta(q^0) \nonumber\\
$$
\begin{eqnarray}
\times [{\bar p}(k^0+M)-E({\bar p})k]\frac{1}{\pi}{\rm
Im}\frac{1}{{q^0}^2-{\vec q}\,^2-m_K^2-\Pi(q^0,{\vec
q}\,)}\biggl|^{q^0=k^0+E({\vec p_N})-\Delta-E({\vec p}\,)}_{{\vec q}\,={\vec k}+{\vec p_N}-{\vec p}} \,,
\label{eq18}
\end{eqnarray}
where we have added the distortion factor for the initial $K^-$ and the
final proton $p$
(exponential factor in the equation), as well as the factor $\theta(q^0)$ of Eq.
(\ref{Eq:cuttheta}).
Taking into account that $\langle \sigma^{\rm tot}_{K^-N} \rangle\simeq 45~{\rm mb}
\simeq \langle \sigma_{pN}\rangle \equiv \sigma$
 for $K^-$-nucleon collisions with $p_{K^-}\simeq 1$ GeV/c and proton-nucleon
 collisions with protons having about $600-700$ MeV kinetic energy, we have implemented a combined
eikonal distortion factor as in \cite{Kishimoto:2007zz,zaki}
\begin{equation}
\int d^3r\to \int d^3r e^{-\int^{\infty}_{-\infty}\sigma\rho(b,z')d z'} \ ,
\end{equation}
where $b$ is the impact parameter $b^2=x^2+y^2$.

The backward differential cross section of the elementary process  $K^- p \to
K^- p$ in the laboratory frame $(d\sigma/d\Omega)_{\rm lab}$ for incoming kaons
of 1 GeV/c is taken to be 8.8 mb/sr, using the $K^- p$ elastic cross-section
data of Ref.~\cite{conforto}. We note that the authors of Ref.~\cite{galself}
take a value of 3.6 mb/sr as an effective way to implement effects of Fermi
motion, Pauli blocking, etc, which here we consider explicitly.

\section{Monte Carlo simulation}

  The procedure outlined above is quite efficient to produce the cross
section for the $(K^-,p)$ reaction, but obviously including only
the quasi-elastic collisions $K^- p \to K^- p$.  There might be other
processes contributing to generate fast protons and, when this is the case,
it becomes advisable to make a simulation of the
reaction. This procedure has been developed in Ref.~\cite{simulation} for the study of
inclusive pionic reactions in nuclei and has also been applied to other
processes, such as photon induced pion and proton emission in nuclei
\cite{rafaloren,rafap}, electron induced proton emission \cite{gil}, nucleon
emission following hypernuclear decay \cite{hyper,garba}, nucleon
emission following kaon absorption in nuclei \cite{Magas:2006fn}, etc.

As sources of fast protons we consider the quasi-elastic
$K^- N$ scattering process, as well as the absorption of the kaon by one and two nucleons. The election of which reaction occurs at a certain point in the nucleus
is done as usual. One chooses a step size $\delta l$ and calculates, by means 
of $\sigma_i \rho \delta l$ with $i={\rm qe, 1N, 2N}$, the probabilities that
any of
the possible reactions happens.
The values of the cross sections are discussed in Sect.~\ref{sec:cross}.
 The size of $\delta l$ is
small enough such that
the sum of probabilities that any reaction occurs is reasonably smaller
than unity. A random number from 0 to 1 is generated and a reaction occurs if
the number falls within the corresponding segment of length given by its probability,
the segments being put
successively in the interval [0-1]. If the random number falls outside the sum of
all segments then
this means that no reaction has been taken place and the kaon is allowed to proceed one further step $\delta l$.

\subsection{Quasi-elastic scattering}

We here describe how the Monte Carlo simulation treats the quasi-elastic
reaction discussed in Sect.~\ref{many-body}. The general strategy is to let the kaon propagate
through the nucleus determining, at each step $\delta l$, whether it can
undergo a quasi-elastic collision, according to the probability $\sigma_{\rm
qe} \rho \delta l$, where $\sigma_{\rm qe}$ is the $K^- N \to K^- N$ elastic cross
section. If there is a quasi-elastic collision at a certain point, then the
initial $K^-$ momentum and the nucleon momentum, randomly chosen within
the Fermi sea, are boosted to their CM frame. The direction of the scattered
momenta is determined according to the experimental cross section. A boost to
the lab frame determines the final kaon and nucleon momenta. The event is kept
as long as the size of the nucleon momentum is larger than the local value of
$k_F$. Since we take into account secondary collisions we consider the reactions
$K^- p \to K^- p$, $K^- p \to K^0 n$ and $K^- n \to K^- n$ with their
corresponding cross sections.

Once primary nucleons are produced they are also followed through the nucleus taking into account the probability that they collide with other nucleons, losing
energy and changing their direction. We follow the procedure detailed in
\cite{simulation,rafap}.

We also follow the rescattered kaon on its way through the nucleus. In the subsequent interaction process we let the kaon experience
whichever reaction of the three that we consider (quasi-elastic, one body absorption, two body absorption) according to
their probabilities.
If the kaon remains after the collision, this procedure continues until it finally emerges
out of the nucleus or it is absorbed by one or two nucleons.

\subsection{One body kaon absorption}
We consider the reactions $K^- N \to \pi \Lambda$ and $K^- N \to \pi \Sigma$,
with all the possible charge combinations. Once again the probability of this
occurring is weighed by their corresponding cross sections and the directions
of the $\pi$ and of the hyperons are also determined in the CM frame.
The system is then
boosted back to the lab frame, where we let the $\Lambda$ or the $\Sigma$ propagate through the nucleus, undergoing quasi-elastic collisions with the nucleons. Once they
leave the nucleus they are allowed to decay weakly into $\pi
N$ providing in this way a source of protons which is not negligible, as we will
see.

\subsection{Two body absorption}

We also take into account the following
processes: $K^- NN \to \Lambda N$ or $K^- NN \to \Sigma N$ with all possible
charge combinations. The probability per unit length for
two nucleon absorption,  $\mu_{K^-NN}$, together with the distribution into the
different possible channels, are discussed in Sect.~\ref{sec:cross}. In these
reactions an energetic nucleon is produced, as well as a $\Lambda$ or a
$\Sigma$ hyperon. Both the nucleon and the hyperon are followed through the nucleus as
discussed above.
Once out of the nucleus, the hyperon is let to decay weakly
into $\pi N$ pairs.
Therefore, the two body absorption process provides a double source of
fast protons, those directly produced in the absorption reactions
and those coming from hyperon decays.

\subsection{Consideration of the $K^-$ optical potential}

We also take into account a kaon optical potential $V_{\rm opt}={\rm Re}\, V_{\rm
opt} + {\rm i}~ {\rm Im}\, V_{\rm opt} $, which will influence the kaon propagation
through the nucleus, especially after a high momentum transfer quasi-elastic
collision when the kaon will acquire a relatively low momentum.

As discussed in the
introduction, one can find in the literature
quite different values for the real part of the potential.
In the present study we vary the strength of the potential,
${\rm Re}\, V_{\rm opt}$, to study the sensitivity of the results: starting from $-60\,
\rho/\rho_0$ \cite{lutz,angelsself,schaffner,galself,Tolos:2006ny}, going
through  $-200\, \rho/\rho_0$ MeV
\cite{gal1,gal2,gal3,gal4,gal,muto,amigo1,amigo2}, and down to $-600\,
\rho/\rho_0$ MeV \cite{akaishi:2002bg,akainew}. For the imaginary part of the
optical potential we take ${\rm Im}\, V_{\rm opt} \approx -60\, \rho/\rho_0$
MeV, as in
the experimental paper \cite{Kishimoto:2007zz} and the theoretical study of
\cite{angelsself}.

In the presence of an optical potential, the kaon spectral function has the
form:
\beq
S_K(\tilde{M}_K )=\frac{1}{\pi} \frac{-2M_K {\rm Im}\,V_{\rm opt}}
{(\tilde{M}^2_K -M_K^2-2 M_K {\rm Re}\,V_{\rm opt})^2 + (2M_K {\rm Im}\,V_{\rm opt})^2}
\,.
\eeq{K_self}
In the Monte Carlo simulation we implement this distribution by
generating a random kaon mass $\tilde{M}_K$ around a central value,
$M_K + {\rm Re}\,V_{\rm opt}$,
which is the bare kaon mass shifted by the real part of the optical potential.
The generated random masses lie within
a certain extension determined by the width of the distribution
$\Gamma_K = -2 {\rm Im}\, V_{\rm opt}$, the size of which is controlled by the
imaginary part of the optical potential. The probability assigned to each value
of $\tilde{M}_K$ follows the Breit-Wigner distribution given by the kaon
spectral function.

\subsection{Final observable}

   After all the processes implemented in the Monte Carlo simulation, some
   particles leave the nucleus, and we select the events
   that contain a fast proton
   in the region of interest. To adapt the calculations
to the experiment of \cite{Kishimoto:2007zz} we keep the protons that emerge
within an angle
of 4.1 degrees in the nuclear rest frame (lab frame). For quasi-elastic
scattering processes this would
correspond to events in which the kaons emerge backwards in the ${\bar K}N$
CM frame and the protons are most energetic, having
of the order of $500-700$ MeV of kinetic energy in
the lab frame.

To facilitate
comparison with experiment, the missing invariant mass of the $^{12}$C$(K^-,p)$
reaction is converted into a binding energy of the kaon, $E_B$, should
the process correspond to the trapping of a kaon in a bound state and emission
of the fast proton, according to
\beq
\sqrt{(E_K+M_{^{12}{\rm C}}-E_p)^2-(\vec{P}_p - \vec{P}_K)^2} = M_{^{11}{\rm B}} + M_{K}
-E_B \,,
\eeq{B_E}
where $E_p,\vec{P}_p$ are the energy and momentum of the observed proton and
$E_K,\vec{P}_K$ are the energy and momentum of the initial kaon.

\subsection{Coincidence simulation}
\label{min_coin}

It is very important to keep in mind that the measurements in the experiment of
\cite{Kishimoto:2007zz} were done in coincidence. The outgoing proton was
measured by the KURAMA spectrometer in the forward direction, while
another detector, the decay counter,
was sandwiching the target. The published spectra was obtained with a
requirement of having an outgoing proton in the KURAMA spectrometer
and at least one charged particle in the
decay counter \cite{Kishimoto_email}.

Obviously, the real simulation of such a coincidence experiment is tremendously
difficult, practically impossible with high accuracy, because it
would require tracing all the charged particles coming out from all
possible scatterings and decays.
Although we are studing many processes and
following many particles in our
Monte Carlo simulation, which is not the case in the Green function method
used in the data analysis \cite{Kishimoto:2007zz},
we can not simulate precisely the real coincidence effect.

What we can do is to eliminate the processes
which, for sure, will not produce a coincidence, a procedure that we 
refer to as minimal coincidence requirement \cite{ISMD2009}. If the kaon in the first
quasi-elastic scattering produces an energetic proton falling into the
peaked region of the spectra, then the emerging kaon will be
scattered backwards.
In our Monte Carlo simulation we can select events were neither the proton
nor the kaon will have any further reaction after such a scattering. In these
cases, although there is a ``good" outgoing proton, there are no charged
particles emerging with the right direction with respect to the beam axis to
hit a decay counter, since the $K^-$ escapes undetected through the backward
direction. Therefore, this type of events must be eliminated for
comparison with the experimental spectra.

As we will see in the next section, the main source of the energetic protons is
$K^-p$ quasi-elastic scattering and, therefore, the minimal coincidence
requirement removes a substantial part of the potentially ``good" events 
changing the form of the final spectrum. Furthermore, events with one or two
nucleon absorption or/and with several quasi-elastic rescatterings have a good
chance of producing a charged particle that goes through the decay counter.
Thus, the final spectrum obtained from our Monte Carlo simulations with minimal
coincidence requirement will probably overshoot the experimental spectrum 
 by an amount which will depend on the capability of the
events having the given energy $E_B$ of producing, apart from the corresponding
energetic proton, additional charged particles hitting the decay counters.

\section{Input cross sections}
\label{sec:cross}

\subsection{\mbox{\boldmath{${\bar K}N$} cross sections}}

The elastic and inelastic two body ${\bar K}N$ cross sections for kaons of about 1 GeV/c are taken from the Particle Data Group (PDG) \cite{PDG}. The values are the following:

$$	\sigma_{K^-p\rightarrow K^-p} =   21.22\ {\rm mb}\,, \quad
	\sigma_{K^-p\rightarrow {\bar K}^0n} =   7.15\ {\rm mb}\,, \quad
	\sigma_{K^-n\rightarrow K^-n} =   18.5\ {\rm mb}$$
$$	\sigma_{K^-p\rightarrow \pi^0 \Lambda} =   4.32\ {\rm mb}\,, \quad	
\sigma_{K^-p\rightarrow \pi^+ \Sigma^-} =   1.76\ {\rm mb}$$
$$	\sigma_{K^-p\rightarrow \pi^- \Sigma^+} =   1.4\ {\rm mb}\,, \quad
	\sigma_{K^-p\rightarrow \pi^0 \Sigma^0} =   1.58\ {\rm mb}$$
$$	\sigma_{K^-n\rightarrow \pi^- \Lambda} =   6.35\ {\rm mb}\,, \quad
	\sigma_{K^-n\rightarrow \pi^- \Sigma^0} =   0.97\ {\rm mb}\,, \quad
	\sigma_{K^-n\rightarrow \pi^0 \Sigma^-} =   1.15\ {\rm mb}$$

From the PDG we also know the total cross-sections:
$$\sigma^{\rm tot}_{K^-p} = 51.7\ {\rm mb}\,,\quad
\sigma^{\rm tot}_{K^-n} = 38\ {\rm mb}\quad \Rightarrow \quad  \langle \sigma^{\rm
tot}_{K^-N}\rangle = 45\ {\rm mb~~~[see\ Eq.\ (\ref{eq18}) ] }\,.$$
Since these are larger than the sum of the partial cross sections that
we are using explicitly, we define:
$$\sigma_{K^- p \rightarrow X} = 14.27\ {\rm mb}\,,\quad
\sigma_{K^- n \rightarrow X} = 10.0\ {\rm mb}\,,$$
which take care about all possible reaction channels,  like $K^- p \to \eta \Lambda$, $K^- p \to \eta \Sigma$ and others, where no fast nucleons come out.
Thus, we introduce an extra segment of length $\sigma_{K^- p,n \rightarrow X}\,
\rho \, \delta l$ in the
interval [0-1] for the Monte Carlo decision of these reactions to occur. If
this is the case, the $K^-$ simply disappears, since the particles produced in these reactions can not contribute to our observable.

\begin{figure}[htb]
\ \ \ \ \ \includegraphics[width=.25\textwidth]{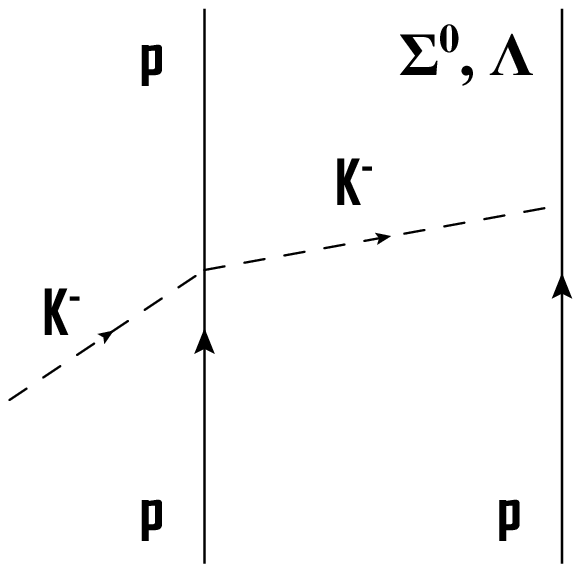}\ \ \ \ \ \ \
\includegraphics[width=.25\textwidth]{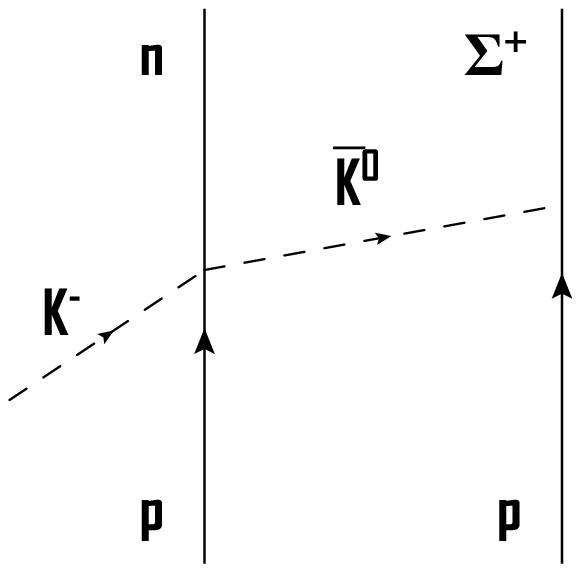}\ \ \ \ \ \ \
\includegraphics[width=.25\textwidth]{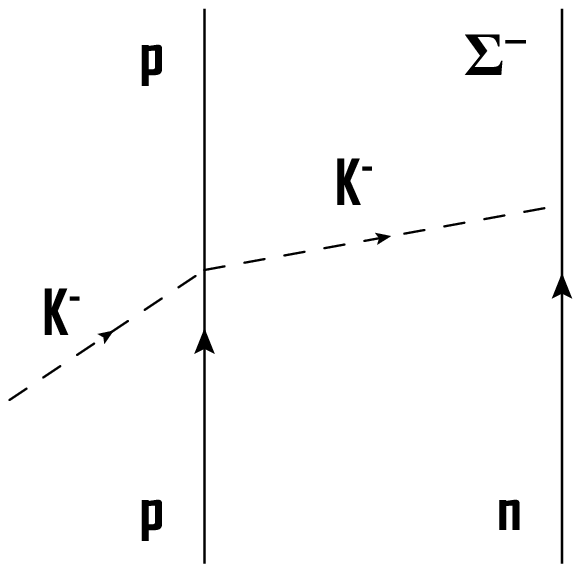}
\vspace{0.5cm}

\noindent
\ \ \ \ \ \includegraphics[width=.25\textwidth]{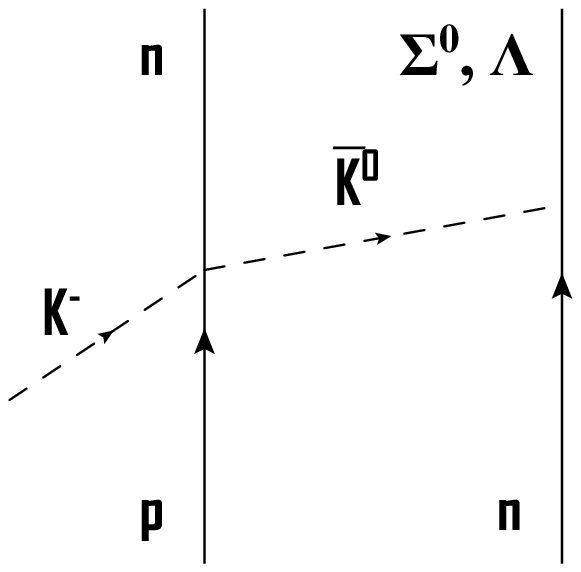}\ \ \ \ \ \ \
\includegraphics[width=.25\textwidth]{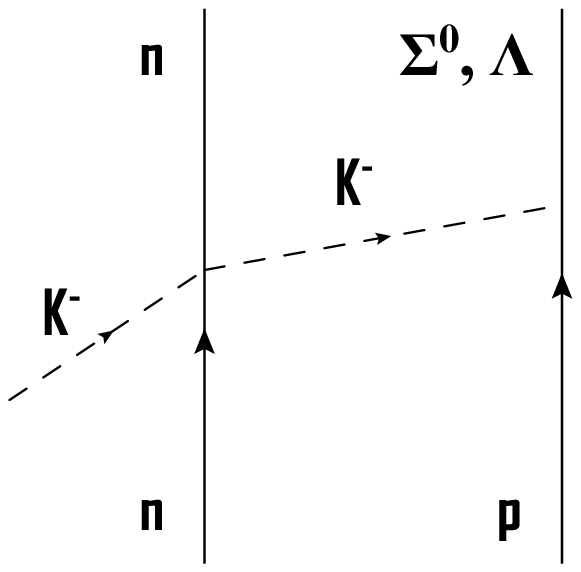}\ \ \ \ \ \ \
\includegraphics[width=.25\textwidth]{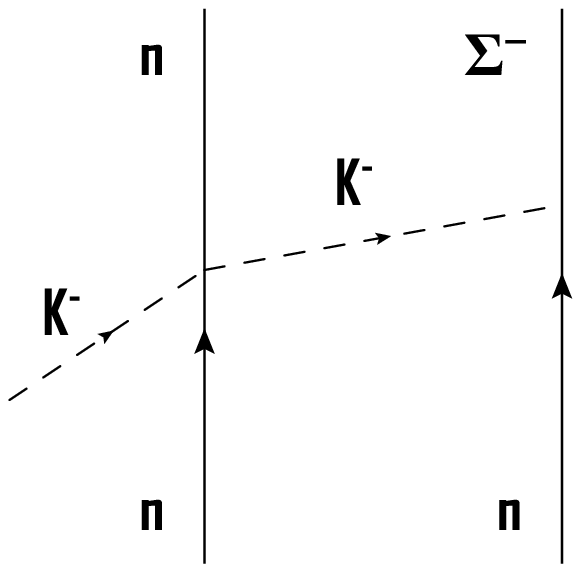}
\caption{The two nucleon $K^-$ absorption diagrams from a kaon-exchange picture.}
\label{fig:3}
\end{figure}

\subsection{Two nucleon absorption cross-sections}

The probability per unit length for two nucleon absorption is proportional to the square of the nucleon density:
$$\mu_{K^-NN}(\rho) = C_{\rm abs} \rho^2\,.$$
We assume a total two body absorption rate of 20\% that of one body absorption at
about nuclear matter density, something that one can infer from
data of $K^-$ absorption in $^4$He  \cite{Katz:1970ng}.
In practice, this is implemented in the following way:
$$ \langle \mu_{K^-NN}\rangle = C_{\rm abs} \langle \rho^2 \rangle =
0.2 \langle \mu_{K^-N} \rangle = 0.2 \langle\sigma^{\rm tot}_{K^-N}\rangle \langle \rho \rangle \,,$$
where $\sigma^{\rm tot}_{K^-N}$ accounts for the total one nucleon absorption cross section and, in symmetric nuclear matter, it is given by:
$$\langle \sigma^{\rm tot}_{K^-N}\rangle=(\sigma^{\rm tot}_{K^-p} + \sigma^{\rm tot}_{K^-n} -
\sigma_{K^-p\rightarrow K^-p} - \sigma_{K^-n\rightarrow K^-n})/2=21.45\ {\rm mb}\,.  $$
Taking $\langle \rho \rangle=\rho_0/2$, where $\rho_0=0.17$ fm$^{-3}$ is normal nuclear matter density, we obtain
$$C_{\rm abs}\approx 6\ {\rm fm}^{5} \,.$$

The different partial processes that can take place in a two nucleon
absorption reaction are:
$$K^- p p \to p \Lambda,\ p \Sigma^0,\ n \Sigma^+ $$
$$K^- p n \to n \Lambda,\ n \Sigma^0,\ p \Sigma^- $$
$$K^- n n \to n \Sigma^- \ .$$

\begin{figure}[htb]
\includegraphics[width=.75\textwidth]{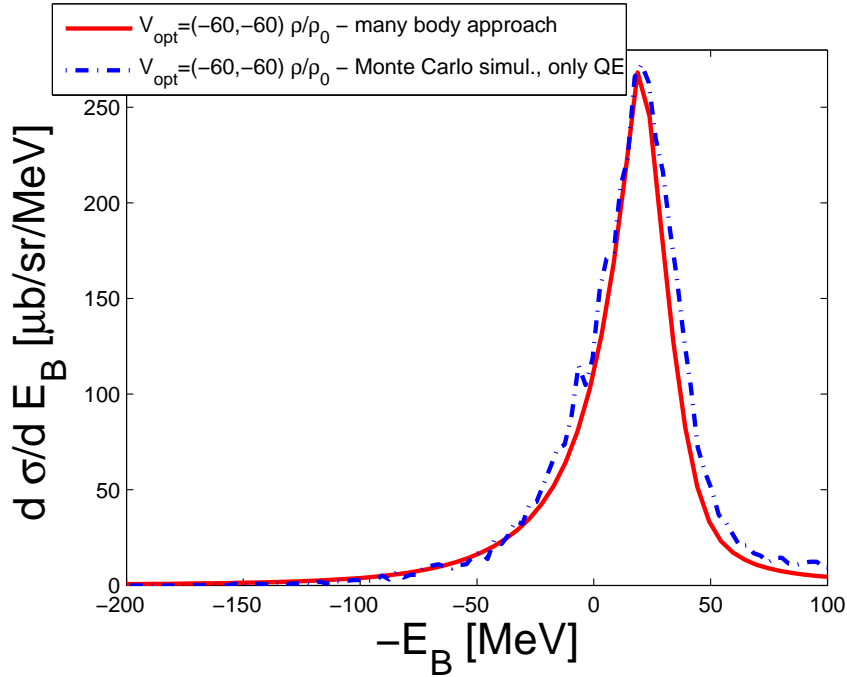}
\caption{The results of the direct many body evaluation (full line), and of
the Monte Carlo simulation considering only the quasi-elastic
scattering processes (dashed-dotted line),
for $V_{\rm opt}=(-60,-60)\rho/\rho_0$ MeV.}
\label{Junko_60_200_A}
\end{figure}

Ideally, their corresponding branching ratios should be obtained from relevant
microscopic mechanisms, such as the kaon-exchange processes depicted in
Fig.~\ref{fig:3}. 
There might be, however, other processes such as, for
instance, those
involving pion exchange. In the present exploratory work, we will
consider a much simpler approach consisting of assigning equal probability to
each of the above reactions. Noting that the chance of the kaon to find a
$pn$ pair is twice as large as that for $pp$ or $nn$ pairs, we
finally assign a probability of 3/10 for having a $p\Sigma$ pair in the final
state of $K^-NN$ absorption, 4/10 for $n \Sigma$, 1/10 for $p \Lambda$ and 2/10 for $n \Lambda$.

\begin{figure}[htb]
\includegraphics[width=.75\textwidth]{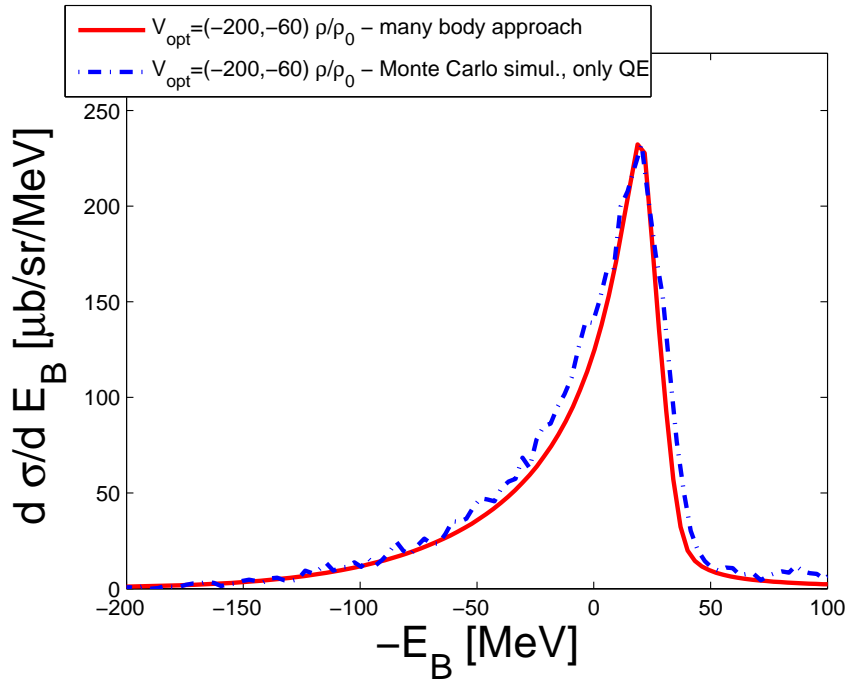}
\caption{The same as Fig. \ref{Junko_60_200_A}, but for $V_{\rm opt}=(-200,-60)\rho/\rho_0$ MeV.}
\label{Junko_60_200_B}
\end{figure}

\subsection{Nucleon and Hyperon cross-sections}

Apart from following the kaons, our calculations also need to consider
the quasi-elastic scattering of nucleons, $\Lambda$  and
$\Sigma$ hyperons on their way through the residual nucleus. The nucleon-nucleon
cross sections $\sigma_{NN}$ for different momenta are taken
from the parameterization of Ref. \cite{NN},
as applied also in other simulations \cite{Magas:2006fn,Magas:2008bp}.
Given the uncertainties in the hyperon-nucleon cross sections, we
may use the relation $\sigma_{YN} = 2\sigma_{NN}/3$ based on a simple
non-strange quark counting rule. This approximation is used for $\Sigma N$ scatterings.
However, other more refined fits to experimental data
also exist and, in the case of $\Lambda N$ scattering, we use the
parameterization of
Ref.~\cite{manolo},  as was also done in Ref.~\cite{Crimea}.

\section{Results and discussion}

In the first place we would like to compare the results obtained with the
diagrammatic many-body method with those of the Monte Carlo simulation when
only quasi-elastic scattering is considered. This is shown in
Figs.~\ref{Junko_60_200_A}, \ref{Junko_60_200_B} for different optical potentials: $V_{\rm
opt}=(-60,-60)\rho/\rho_0$ MeV  and  $V_{\rm
opt}=(-200,-60)\rho/\rho_0$ MeV correspondingly. As we can see, the two
calculations are practically identical in the region of interest. The Monte
Carlo simulation produces slightly larger cross sections because it also takes
into account the multiple quasi-elastic scattering processes. It is clear that
these additional events, which contain a 
``good" final proton and more than one quasi-elastic collision,
are rather rare. However, as we will
see, kaon absorption mechanisms produce a substantial
amount of energetic nucleons which need to be taken into account.

\begin{figure}[htb]
\includegraphics[width=.75\textwidth]{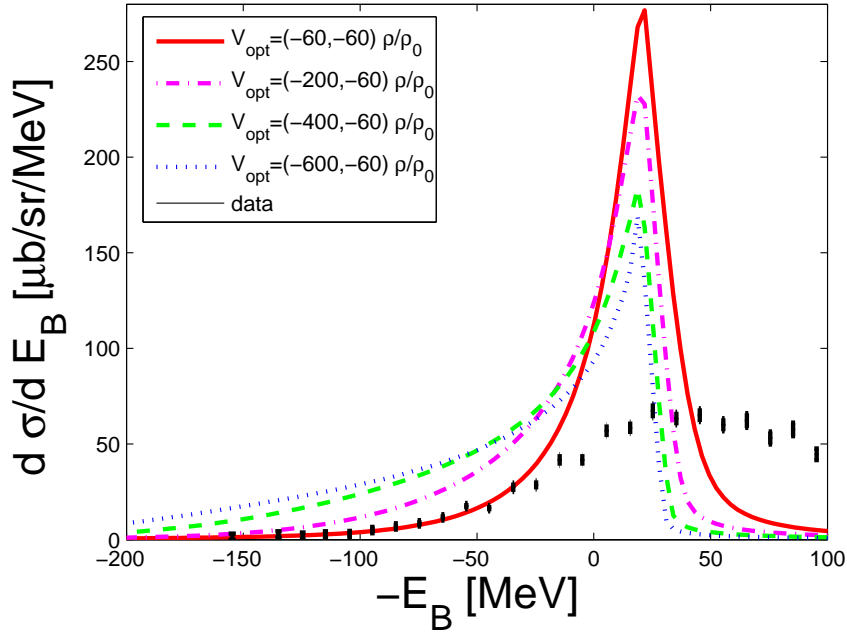}
\caption{Results obtained using the many body method for kaon potential depths
of $60$ MeV, $200$ MeV, $400$ MeV and $600$ MeV at normal nuclear density. 
Experimental data are shown with black bars.}
\label{fig23a}
\end{figure}

\begin{figure}[htb]
\includegraphics[width=.75\textwidth]{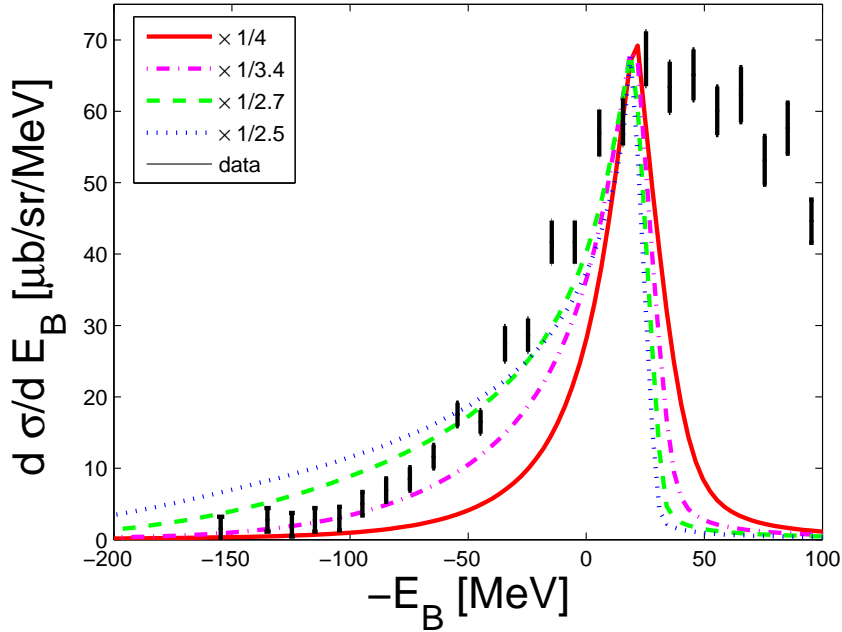}
\caption{The same results as in Fig. \ref{fig23a},
but rescaled to the heigth of the experimental spectrum
\cite{Kishimoto:2007zz}. Experimental data are shown with black bars.}
\label{fig23b}
\end{figure}

Before showing the contributions of the new processes, let us first explore the
sensitivity of the spectrum to the kaon optical potential.  
In Figs.~\ref{fig23a},~\ref{fig23b} we show the results obtained with the many body method 
employing potential
depths of $60$ MeV, $200$ MeV, $400$ MeV and $600$ MeV at normal nuclear 
density. In Fig. \ref{fig23a} we see the absolute distributions
plotted together with the experimental spectrum \cite{Kishimoto:2007zz}.  
Increasing the
depth of the kaon optical potential produces an enhancement of the
cross section in the bound region of kaons, as one might expect intuitively.
The height of the theoretical distributions is much larger than
that of the experimental cross section. We
have tested that our theoretical normalization is correct.
Indeed, if we remove the distortion of the incoming kaons and ougoing protons
in the many body method, we obtain that the strength of the integrated cross
section for the reaction on $^{12}$C
is six times the one of the elementary reaction, $K^- p
\to K^- p$, at backward angles.
The distortion implemented here reduce the nuclear cross section by 
about a factor $3.5$,
which is also the same distortion effect obtained in \cite{zaki}.

The different size of the theoretical distribution as compared to the
experimental data is in fact showing the removal of events implemented by the
concidence test applied in Ref.~\cite{Kishimoto:2007zz}, demanding that some
extra charged particle is detected in a decay counter surrounding the target
together with the forward fast proton. It is, however, claimed in
\cite{Kishimoto:2007zz}  that the required coincidence does not change the
shape of the spectrum. Assuming this, we can rescale our calculations to give
them the size of the experimental distribution, as is illustrated in Fig.~\ref{fig23b}. 
The region of main interest corrresponds to a deep
binding energy for the kaon (i.e. high momenta for the proton) of about 50 MeV
or more to the left of the peak,  since the quasi-elastic approach ignores many
processes populating the spectrum at low proton momenta. We observe that, 
to obtain a good description of the spectrum from $-E_B \sim 0$
down to $-E_B$ of around $-100$ MeV one would need optical
potential depths as large as $400$ MeV, or even $600$ MeV, at normal nuclear
density, not the $190$ MeV claimed in Ref.~ \cite{Kishimoto:2007zz}, and even
then we observe that the shape is not well reproduced for any of the
potentials.

\begin{figure}[htb]
\includegraphics[width=.75\textwidth]{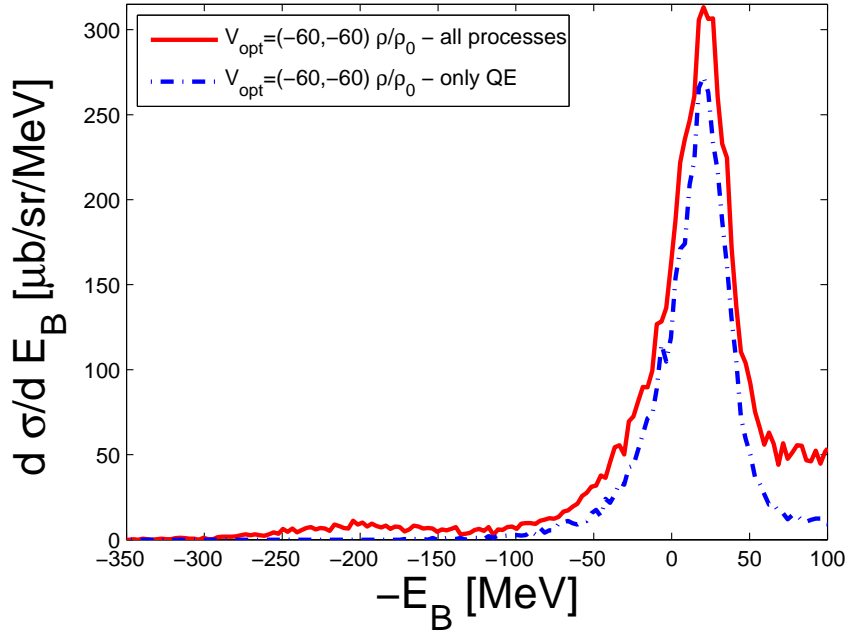}
\caption{Calculated proton spectra
 with  $V_{\rm opt}=(-60,-60)\rho/\rho_0$ MeV,
  taking into account only quasi-elastic processes (dash-dotted line),
  and including all processes (solid line).}
\label{fig1}
\end{figure}

\begin{figure}[htb]
\includegraphics[width=0.85\textwidth]{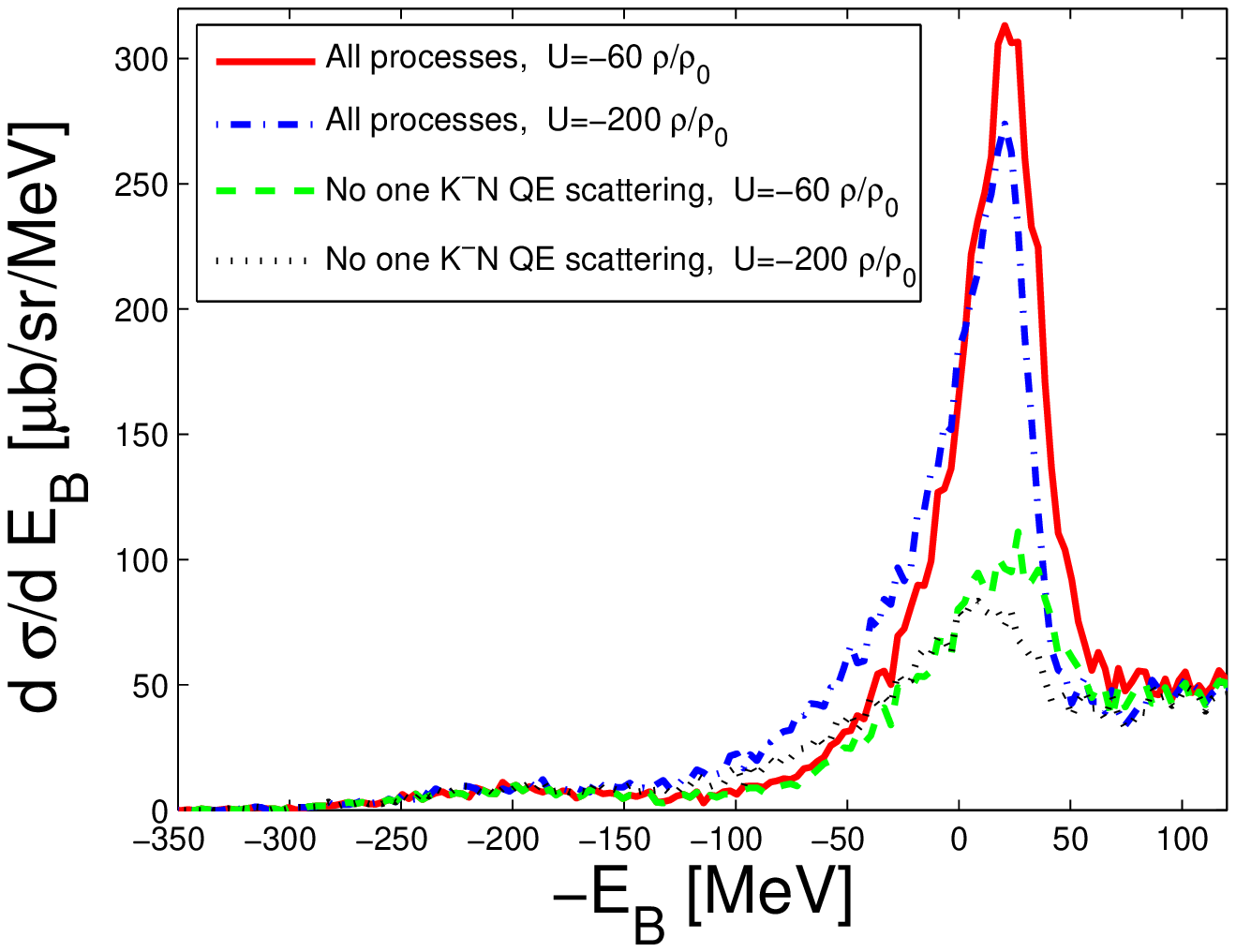}
\caption{Calculated $ ^{12}{\rm C}(K^-,p)$ spectra
for $V_{\rm opt}=(-60,-60)\rho/\rho_0$ MeV and  $V_{\rm opt}=(-200,-60)\rho/\rho_0$ MeV
  taking into account all contributing processes (solid and dot-dashed lines), 
  and imposing the minimal
  coincidence requirement (dashed and dotted lines).}
\label{fig2}
\end{figure}

In our opinion, these results indicate the existence of other contributions
from processes that are not yet taken into account and/or that the assumption
of an energy independent reduction factor due to the coincidence requirement
might not be correct.
These effects can be investigated within the Monte
Carlo simulation developed in this work. In Fig.~\ref{fig1}
we show  the results
of the Monte Carlo simulation obtained with an optical potential
$V_{\rm opt}=(-60,-60)\rho/\rho_0$ MeV,  taking into account only quasi-elastic
processes (dash-dotted line) and considering as well one nucleon and two
nucleon absorption processes (solid line).
We can see that there is a non-negligible amount of
strength gained in the region of ``bound kaons" due to the new mechanisms.
Although not shown separately in the figure, we have seen
that one nucleon absorption and multi-scatterings
contribute to the region $-E_B > -50$ MeV. To some extent, this strength
can be simulated by the parametric background used in
\cite{Kishimoto:2007zz}. However, this is not true anymore for the two 
nucleon absorption processes,
which contribute to all values of $-E_B$, starting from almost as low as 
$-300$ MeV.

\begin{figure}[htb]
\includegraphics[width=.75\textwidth]{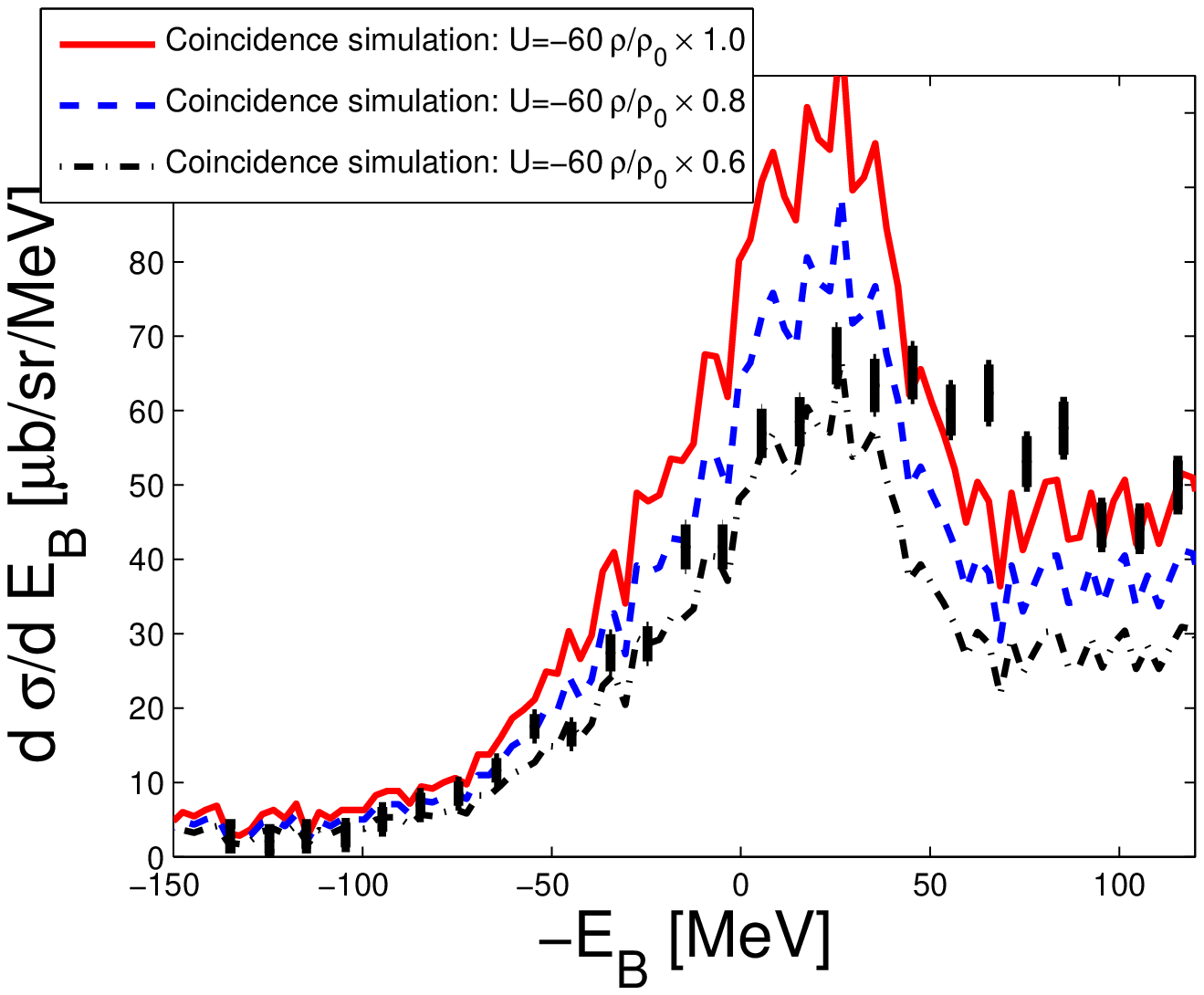}
\caption{The $ ^{12}{\rm C}(K^-,p)$ spectrum
 obtained with $V_{\rm opt}=(-60,-60)\rho/\rho_0$ MeV and
 the minimal coincidence requirement, for several 
 reduction factors. Experimental points are taken 
 from \cite{Kishimoto:2007zz}.}
\label{fig3a}
\end{figure}

\begin{figure}[htb]
\includegraphics[width=.75\textwidth]{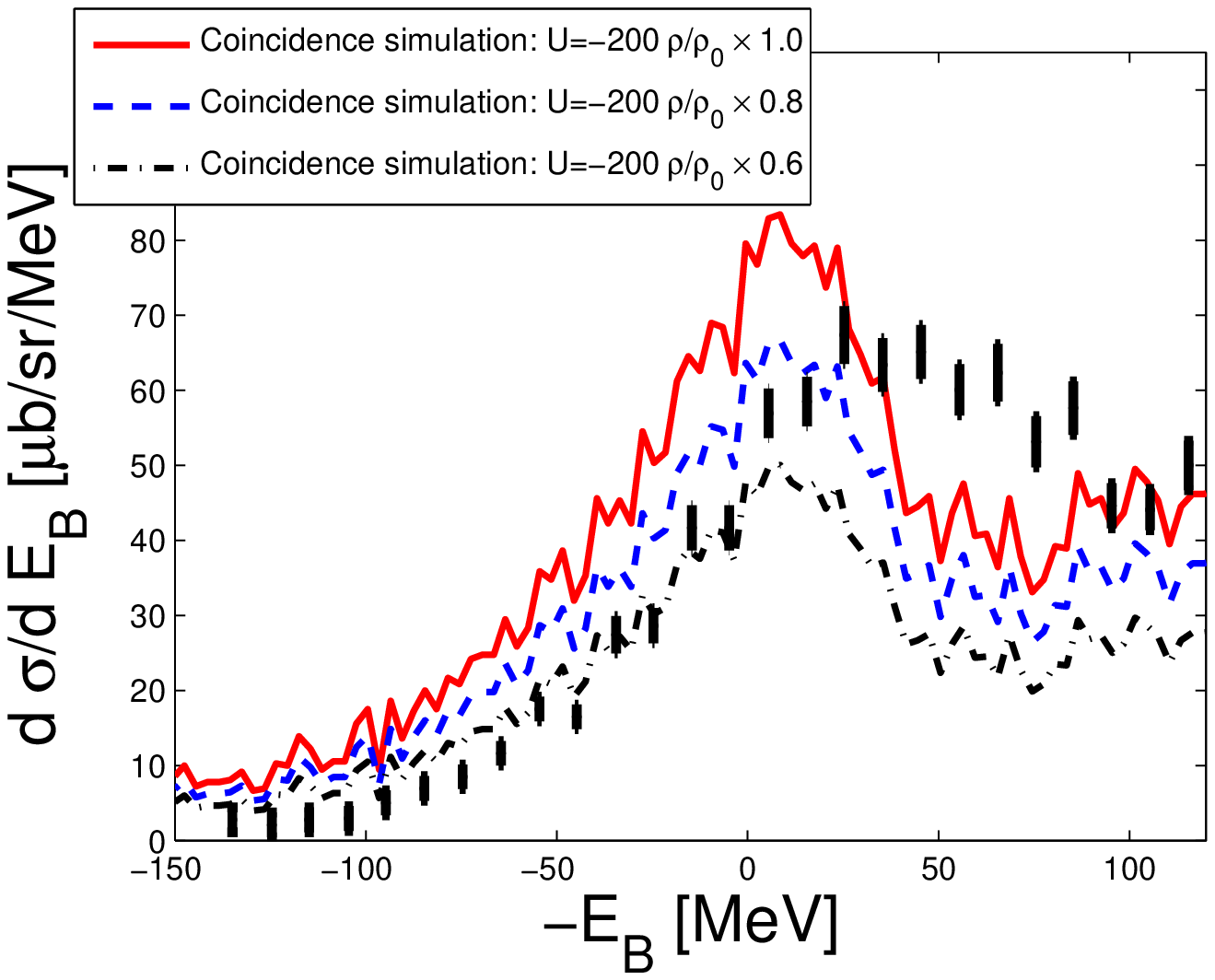}
\caption{The same as Fig. \protect\ref{fig3a}, but  for
 $V_{\rm opt}=(-200,-60)\rho/\rho_0$ MeV.}
\label{fig3b}
\end{figure}

It is very important to keep in mind that in the spectrum of \cite{Kishimoto:2007zz}
the outgoing forward protons were measured in coincidence
with at least one charged particle in the decay counters surrounding the target.
While a detailed simulation of these experimental conditions is prohibitive we
can at least see their consequences by applying the minimal
coincidence requirement. As described in Sect.~\ref{min_coin}  we eliminate the
events that, for sure, will not produce a coincidence, i.e. those
in which, after a primary quasi-elastic collision producing a fast forward 
proton and a backward kaon, neither particle suffer any further
reaction. While it is
clear from Fig.~\ref{fig1} that the main source of 
energetic protons in the $ ^{12}{\rm C}(K^-,p)$ spectrum is the
$K^-p$ quasi-elastic scattering process, many of these potentially 
``good" events will be eliminated by the
the minimal coincidence requirement. As a result, the shape of the spectrum will
change substantially, as clearly illustrated in 
Fig.~\ref{fig2} upon comparing the bare spectrum obtained with a kaon potential
depth of 60 MeV (solid line) with that
obtained after the minimal coincidence cut (dashed line). The figure also shows
the spectra corresponding to a potential depth of 200 MeV, before (dot-dashed line) and
after the coincidence cut (dotted line). We clearly see that 
the sensitivity of the spectra in the bound region to the optical 
potential employed 
is practically lost when the coincidence requirement is applied. These
results demonstrate the limited capability of the $(K^-,p)$ reaction with
in-flight kaons to infer the depth of the kaon optical potential. Actually,
the bare spectrum would be a more appropriate observable for this task.

To further understand the effects of the coincidence requirement we introduce
additional constant suppression factors to the calculated spectrum
\cite{ISMD2009}, as seen in Figs.~\ref{fig3a} and \ref{fig3b}. Figure~\ref{fig3a}
shows our results using a shallow kaon nucleus optical potential, $V_{\rm
opt}=(-60,-60)\rho/\rho_0$ MeV, as obtained in chiral models. Comparing to
experimental data, we can conclude that our results would need a reduction
factor of about $\sim 0.7$, more or less homogeneous in the ``bound" region,
$-E_B < 0$ MeV, while the suppression should be weaker in the continuum, and
basically negligible for $-E_B > 50$ MeV. This picture is natural from the
physical point of view, because the spectrum to the right of the peak is
populated with lower momentum protons. These are mostly produced in many
particle final states, which have a better chance to score the coincidence
detectors.

However, if we look at Fig.~\ref{fig3b},
where the
calculations with  a deep kaon nucleus optical potential of  $V_{\rm
opt}=(-200,-60)\rho/\rho_0$ MeV are shown, we can conclude that it is
much more
difficult to obtain an overall description of the data with such a
potential,
even admitting a strong supression in the bound region and a negligible
one in
the continuum.

In spite of the above described behavior,
one cannot conclude that the experimental
spectrum supports especially one potential depth over
the other.
However, we want to make clear that, in
trying to reproduce the actual data, one necessarily introduces large
uncertanties due to the experimental set up. Contrary to what it is assumed in
Ref.~\cite{Kishimoto:2007zz}, we have clearly seen in Fig.~\ref{fig2}, that the
spectrum shape is affected by the required coincidence. In fact, the distorsion
of the experimental spectrum due to the coincidence requirement can easily be
much bigger than the difference between different potential depths, as seen by
the sensitivity of the spectrum to the optical potential displayed in
Figs.~\ref{fig23a} and \ref{fig2}. Thus, the experiment of
Ref.~\cite{Kishimoto:2007zz} is not appropriate for extracting information on
the kaon optical potential. The theoretical analysis of \cite{Kishimoto:2007zz}
was based on the assumption that the shape of the spectrum does not change with
the coincidence requirement. Since we have shown this not to be case,  the
conclusions obtained there do not hold. Certainly, the experimental data 
without the coincidence requirement of \cite{Kishimoto:2007zz} would be a much 
more useful observable.

\section*{Acknowledgments}

This work is partly supported by
the contracts FIS2006-03438, FIS2008-01661 from MICINN
(Spain), by CSIC and JSPS under the Spain-Japan research Cooperative program,
 and by the Ge\-ne\-ra\-li\-tat de Catalunya contract 2009SGR-1289. We
acknowledge the support of the European Community-Research Infrastructure
Integrating Activity ``Study of Strongly Interacting Matter'' (HadronPhysics2,
Grant Agreement n. 227431) under the Seventh Framework Programme of EU.
J.Y. is a Yukawa Fellow and this work is partially supported by the
Yukawa Memorial Foundation.

\end{document}